\documentclass[conference]{IEEEtran}

\usepackage{cite}
\ifCLASSINFOpdf
 \usepackage[pdftex]{graphicx}
\else
\fi

\usepackage[cmex10]{amsmath}

\usepackage{url}
\usepackage{hyperref}
\usepackage{multirow, tabularx}
\usepackage{array}

\newcommand{\slfrac}[2]{\left.#1\middle/#2\right.}

\newcommand{\enquote}[1]{``#1''} 
\usepackage[usenames, dvipsnames]{color}

\newif\ifblind

\blindtrue
\blindfalse

\IEEEoverridecommandlockouts

\usepackage{tikz}
\newcommand\copyrighttext{%
  \footnotesize Copyright \textcopyright 2017 IEEE. Personal use of this material is permitted.
  Permission from IEEE must be obtained for all other uses, in any current or future 
  media, including reprinting/republishing this material for advertising or promotional 
  purposes, creating new collective works, for resale or redistribution to servers or 
  lists, or reuse of any copyrighted component of this work in other works. 
  DOI: \href{https://doi.org/10.1109/QRS.2017.51}{10.1109/QRS.2017.51}
  }
\newcommand\copyrightnotice{%
    \begin{tikzpicture}[remember picture,overlay]
    \node[anchor=south,yshift=10pt] at (current page.south) {\fbox{\parbox{\dimexpr\textwidth-\fboxsep-\fboxrule\relax}{\copyrighttext}}};
    \end{tikzpicture}%
}

\begin{document}
%
\title{Should I Bug You? Identifying Domain Experts in Software Projects Using Code Complexity Metrics}

\ifblind
    \author{\IEEEauthorblockN{redacted for peer review}
    \IEEEauthorblockA{XX, XX\\
    XX\\
    XX, XX\\
    {\{xx.xx\}}@example.org}
    }
\else
    \author{\IEEEauthorblockN{Ralf Teusner, Christoph Matthies, Philipp Giese}
    \IEEEauthorblockA{Hasso Plattner Institute, University of Potsdam\\
    August-Bebel-Str. 88\\
    Potsdam, Germany\\
    {\{firstname.lastname\}}@hpi.de}
    }
\fi

\maketitle

\copyrightnotice

\begin{abstract}
In any sufficiently complex software system there are experts, having a deeper understanding of parts of the system than others.
However, it is not always clear who these experts are and which particular parts of the system they can provide help with.
We propose a framework to elicit the expertise of developers and recommend experts by analyzing complexity measures over time.
Furthermore, teams can detect those parts of the software for which currently no, or only few experts exist and take preventive actions to keep the collective code knowledge and ownership high.
We employed the developed approach at a medium-sized company.
The results were evaluated with a survey, comparing the perceived and the computed expertise of developers.
We show that aggregated code metrics can be used to identify experts for different software components.
The identified experts were rated as acceptable candidates by developers in over 90\% of all cases.
\end{abstract}

\begin{IEEEkeywords}
domain experts; expert identification; software metrics; software quality;
\end{IEEEkeywords}

\section{Introduction}
The \emph{Bus Number} was informally defined by Coplien as the amount of developers that \enquote{would have to be hit by a truck (or quit) before the project is incapacitated}~\cite{williams_pair_2002}, with the worst answer to the question being \enquote{one}.
Losing developers in a software projects is especially disruptive if they were experts for a part of the system and contributed to the bus number.
The most common ownership model for code is that of \emph{subsystem ownership}~\cite{nordberg03}, in which an expert takes primary responsibility for one or more software components.
However, in large and usually also distributed software projects it is often not clear who these experts are and which parts of the system they have deep knowledge in~\cite{Ehrlich:2006:LEG:1170744.1171374, Faraj:2000:CES:970301.970322, herbsleb1999beyondconwayslaw}.
As such, when needing detailed knowledge of a subsystem, the expert needs to be found in a time-consuming, manual process, including possibly multiple referrals.
Management has a vested interested in determining and possibly increasing the bus number, i.e. the amount of expert developers, in order to make the project more resilient.
Uneven distributions of experts can additionally be an indicator of low collective code ownership, a core concept of Extreme Programming.
It describes the idea of every programmer being able to improve any code anywhere in the system~\cite{Beck1999}.
High levels of collective code ownership can help in ensuring that the overall design is based on technical decisions, rather than following Conway's Law\footnote{Conway's Law states that the software structure developed in organizations reflects the organizational communication structure.}~\cite{conway}. It thus helps to encourage developers to feel more responsible for the quality of the whole project~\cite{nordberg03, agilealliance15}.
The Analyzr framework enables analyses on the expertise of developers for parts of the system based on proven code complexity measures.
It is publicly available as open-source software on GitHub
\ifblind
    \footnote{URL redacted for peer review}
\else
    \footnote{\url{https://github.com/frontendphil/analyzr}}
\fi
under the MIT license.

\section{Related Work}
Related work for the identification of domain experts using code complexity measures can be found mainly in the areas of measuring and aggregating source code metrics, code ownership as well as alternative methods of expert identification techniques.

\subsection{Source Code Metrics}
Coleman et al.~\cite{coleman1994using} evaluate different software metrics in regards to their suitability as software maintainability predictors.
The authors settle on the McCabe complexity as well as a set of Halstead’s metrics.
They point out that their approach could help \enquote{maintainers guide their efforts}.
Clark et al.~\cite{clark2008measuring} explore the use of software metrics in the area of autonomous vehicles.
The authors rely solely on the McCabe complexity as they point out that a correlation between code errors and a high  complexity was found~\cite{watson1996structured}.
Nagappan et al.~\cite{nagappan2006mining} employ the  complexity alongside a set of object oriented metrics to ascertain software properties.
They argue that there is no universal set of metrics and that metrics have to be chosen on a per-project basis.

\subsection{Aggregation of Code Metrics}
Vasilescu et al.~\cite{vasilescu2011no, vasilescu2011you} point out the need for aggregating metrics as most of them are defined on a micro level, such as functions or classes.
However, conclusions have be drawn on a component or system level.
Mordal et al.~\cite{mordal2012software} point out the deficiencies of using the average to aggregate metric scores.
They introduce the \emph{Squale} model, which allows effectively comparing different metric values by normalising them into a given interval of values.

\subsection{Code Ownership}
Code ownership describes the approach of assigning source code or entire software systems to their human owners.
these models can range all the way from a single \emph{product specialist}, managing all the code to \emph{collective ownership}, where responsibilities are shared amongst all developers~\cite{nordberg03}.

Avelino et al.~\cite{Avelino2016} propose an automated approach to estimating the \emph{Truck Factor} (TF) of a project, a measure in the agile community of how prepared for developer turnover a project is.
The authors state that the majority (65\%) of the 133 surveyed systems extracted from GitHub had a TF $ \leq$ 2.

Bird et al.~\cite{Bird2011} show that high levels of ownership were associated with less defects in the context of the two software products Windows Vista and Windows 7.
Foucault et al.~\cite{Foucault2014} attempted to replicate Bird's study in the context of free/libre and open-source software projects (FLOSS).
They explored the relationship between ownership metrics and module faults in seven FLOSS projects, but only found a weak correlation.
Thus, the authors conclude that the results of ownership studies performed using closed-source projects, which showed ownership metrics as accurate indicators of software quality, do not generalize to FLOSS projects.

Thongtanunam et al.~\cite{Thongtanunam2016} suggest complementing code ownership heuristics that rely on file authorship with code review metrics.
This also includes those developers that contributed to the code by critiquing changes and suggesting edits.

\subsection{Expert Identification}
McDonald and Ackermann~\cite{McDonald2000} describe a general architecture for \emph{expertise locating systems}.
They point out that these systems are not designed to replace key operational roles, such as a senior employee or guru, but can decrease workload and support decisions where previously no help was available.
Furthermore, the authors state that organizationally relevant sources of information and heuristics need to be fitted to the work setting.
They conclude that recommendation systems can help in finding experts who may not otherwise have been identified.

Schuler et al.~\cite{schuler2008mining} present an approach for retrieving the expertise of developers through analysis of method changes as well as method calls based on data gathered from code repositories.
However, the authors do not take into account metrics that would indicate the quality of the code being examined.

Anvik et al.~\cite{Anvik:2006} present an approach to recommending a set of developers suited for assignment to bug tickets in the context of the \emph{Mozilla} and \emph{GCC} projects.
They employed support vector machine classifiers, based on the one-line summary and full text description of collected bug reports that developers had previously been assigned to or had resolved.
The feature vector was based on the frequency of terms in the text.
The authors claim a precision of 64\% precision for the \emph{Firefox} project.

In the same problem domain, Tian et al.~\cite{Tian2016} propose a model for assigning developers to bug reports.
Their model combines activity-based (developers who fixed similar bugs in the past) as well as location-based techniques.
The authors report that the most important similarity feature in their unified model was whether a developer had previously edited a file containing a potential bug.
The proposed framework expands on this idea by enabling the rating of changes based on complexity metrics.

Venkataramani et al.~\cite{Venkataramani:2013} built a model of developer expertise in a target domain by mining developers activities in different open source projects.
The example used is a recommendation system for StackOverflow based on data mined from GitHub.
The system is based on \emph{author/technical term} mappings extracted from source code and commits by authors in a bag-of-words model.
In a subsequent step, technical terms associated with autors are mapped to StackOverflow tags.
Unfortunately. the details on this mapping are not presented.
The authors state that for a sample of 15 developers, 7 of them answered StackOverflow questions with tags that the model had discovered she was proficient in.

LaTozza et al.~\cite{latoza2006maintaining} point out that expertise not only means knowing more than others but also knowing where to look for the answer, or whom to ask.
Their research revealed that interruptions by colleagues were ranked second when it came to issues hindering developers from working.
Therefore approaches that identify more specific component experts and thereby spread the workload, would alleviate this burden for current experts.

\section{Complexity Measures}
\label{sec:complexitymeasures}
The selection of appropriate complexity measures to determine developer expertise is vital to the quality of Analyzr results.
Code metrics have to be chosen on a case by case basis as no single set of metrics can fit all use cases and contexts.
As a basis for selecting metrics we propose Kaner's \enquote{Ten Measurement Factors}~\cite{hoffman2000darker} for software metrics as well as further literature concerning the most relevant metrics for software design~\cite{riaz2009systematic, german2003software, kaner2004software, basili1996validation}.
We  employed metrics that measure independent aspects, used different approaches and incorporated different code parts in order to compute their results~\cite{graves2000predicting}.

\begin{figure}
    \centering
    \includegraphics[width=\columnwidth]{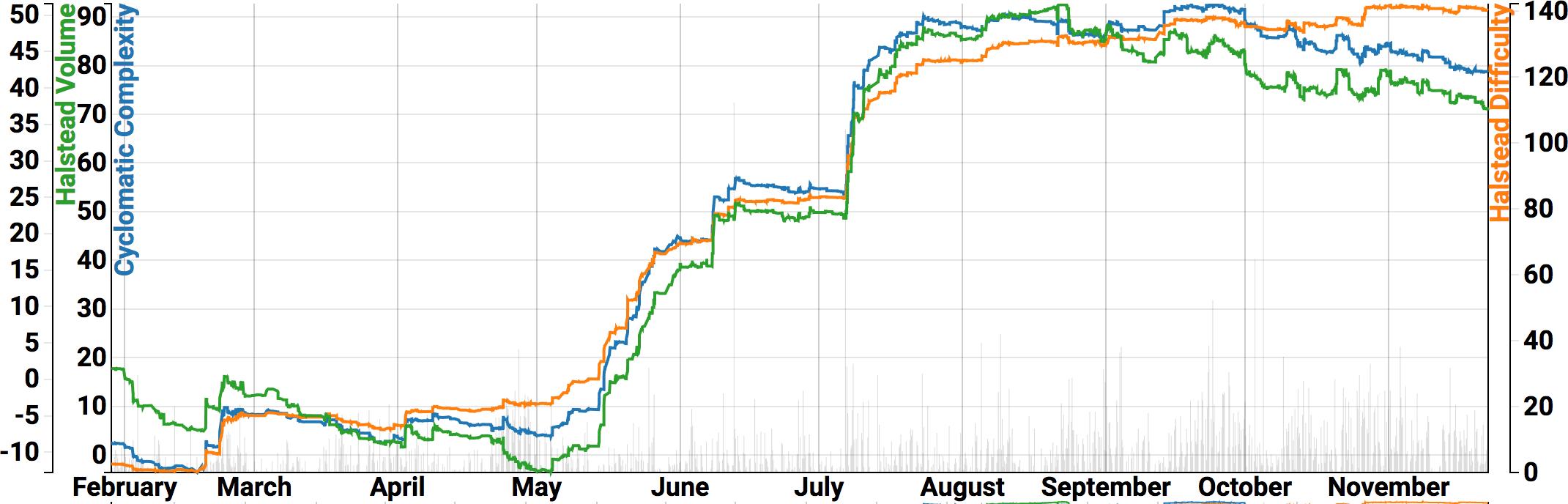}
    \caption{Excerpt of changes in Halstead Volume (green) and Difficulty (orange) as well as Cyclomatic Complexity (blue) in the back end of the studied company. Deltas were oriented to indicate improved code quality, e.g. lower complexity, with rising chart lines.
    For brevity all charts were combined, however, as the deltas are from different domains, the absolute values do not allow direct comparisons.
    The grey bars indicate the amount of commits.}
    \label{fig:metrics}
\end{figure}

Figure~\ref{fig:metrics} shows a summary of the development of a selection of code metrics at the company under study.
In the shown timeframe, the company transitioned from a \enquote{start-up} phase, where the focus lay on fast feature introduction to support first customers, to a \enquote{sustainable} phase, focusing on a maintainable code base.
The shift of focus is apparent in the increase of code quality around the end of 2011, where several refactorings took place.
Since then, the Cyclomatic Complexity as well as the Halstead Volume have slowly begun to degrade again, as new code was added.
This shows that real world circumstances are directly reflected in source code metrics, allowing insights into the development process.

In our study, the following three code complexity measurements were employed:

\subsection{McCabe Complexity}
The McCabe or \emph{cyclomatic} complexity measure is derived from the amount of possible control flows that exist in a program~\cite{mccabe1976complexity}.
A low McCabe complexity would thus be computed for a method with no branches and or one with only a simple type check in them. 
The metric operates on an abstract syntax tree and does not rely on a specific programming language~\cite{watson1996structured}.
The cyclomatic complexity $v$ of a program $G$ is defined as:
\begin{equation*}
    v(G) = e - n + 2p
\end{equation*}
where $e$ is the number of edges, $n$ is the number of nodes, and $p$ is number of exit nodes (amount of possible program flow exits).

\subsection{Halstead Metrics}
Halstead introduced a set of metrics based on the concept that the complexity of a program will increase with the addition of new operators and operands~\cite{halstead1977potential}.
Halstead based his metrics on four key variables: $\eta_1$, $N_1$, the number of distinct and total operators, as well as $\eta_2$, $N_2$ the number of distinct and total operands.
We focus on \emph{difficulty} and \emph{volume}, as these are the most accepted in literature~\cite{shen1983software, coleman1994using}. They are defined as:

\begin{itemize}
    \item \emph{Volume}: $V = N \times \log_2 \eta$
    
    \vspace{0.2cm}
    
    \item \emph{Difficulty}: $D = \displaystyle \frac{\eta_1}{2} \times \frac{N_2}{\eta_2}$
\end{itemize}

\subsection{Coupling}
Coupling metrics describe the amount of dependencies between classes in a package and those outside of it.
Software is easier to understand and maintain if it is split into modules of reasonable size~\cite{henry1981software}.
However, coupling metrics are only applicable to programming languages that support object-oriented idioms such as classes and imports.
While this is the case with Java, JavaScript does not natively support these concepts~\cite{crockford2008javascript}.
There are two types of coupling:
\begin{itemize}
    \item \emph{Efferent coupling}, also known as Fan-Out,
    \item \emph{Afferent coupling}, also known as Fan-In.
\end{itemize}
Efferent coupling describes the number of classes that a given class depends on.
Therefore the class can be affected by changes made elsewhere. 
While a high value in this metric does not necessarily represent bad design, it is often an indicator that the class has too many responsibilities, is poor in maintainability and should be split~\cite{redin2008use}.
Afferent coupling describes the amount of classes that depend on a given class.
Changes in that class will affect all classes which depend on it.
High values in either case of these metrics can be indicators for problems.
Classes which have high values for both efferent and afferent coupling are often a source of bugs \cite{nagappan2006mining}.

\section{The Analyzr Framework}
\label{sec:analyrframework}
\begin{figure}
    \centering
    \includegraphics[width=\columnwidth]{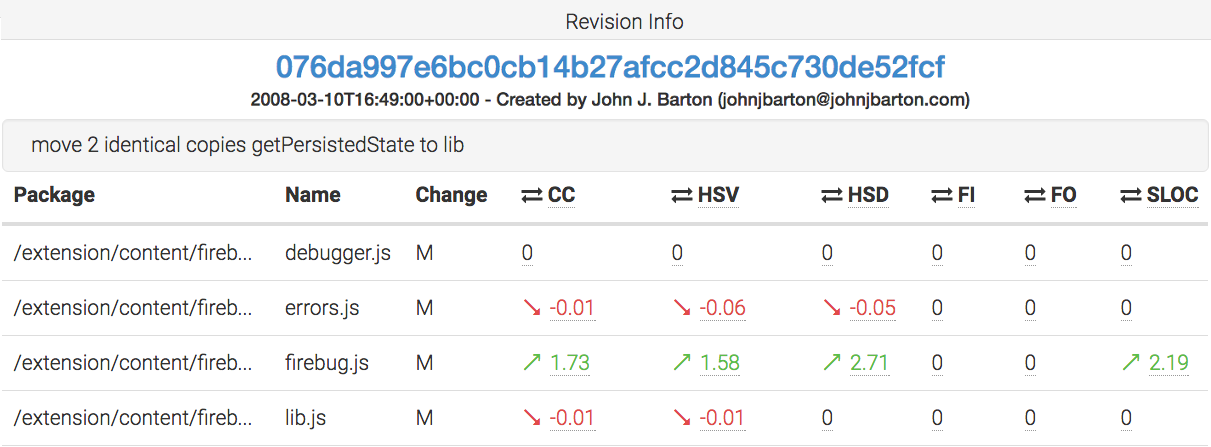}
    \caption{Analyzr screenshot showing the changes in complexity measures for a single commit of the \emph{Firebug} project.}
    \label{fig:commit}
\end{figure}

\begin{figure}
    \centering
    \includegraphics[width=0.8\columnwidth]{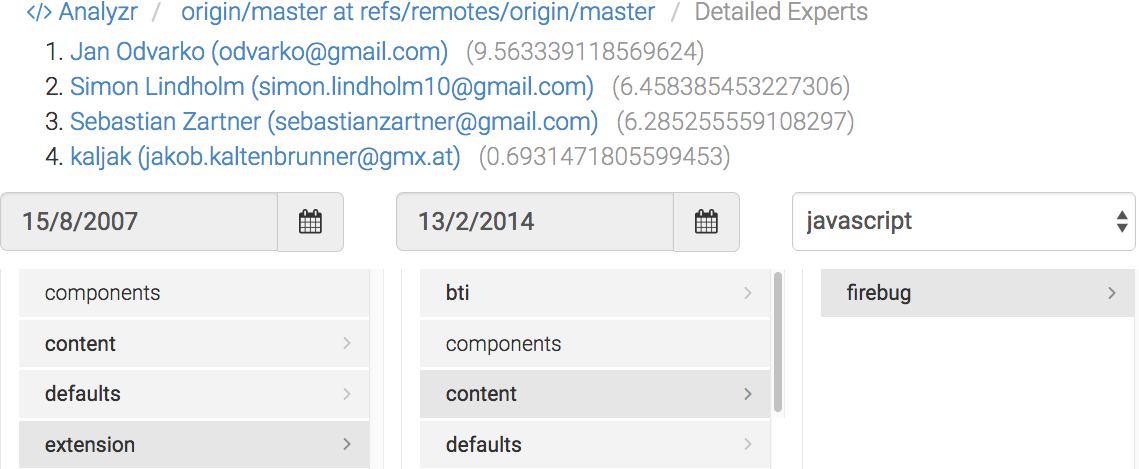}
    \caption{Analyzr screenshot showing developers ranked by expertise for a component of the \emph{Firebug} package.}
    \label{fig:expertise}
\end{figure}

Analyzr identifies component domain experts by aggregating the results of various code complexity measurements on collected development data.
Figure~\ref{fig:commit} shows the changes in metrics for a single commit\footnote{\url{https://github.com/firebug/firebug/commit/076da997e6bc0cb14b27afcc2d845c730de52fcf}}.
Using this data, component domain experts are identified, see Figure~\ref{fig:expertise}.

Analyzr abstracts from the different version control systems that an organisation uses in order to allow analyses on a unified view of the repositories.
For every repository, commit information, such as modified files and author, as well as meta-information about the repository itself, such as a list of branches, are gathered.
This collected data is then used as input for proven third-party tools, specialized for the employed programming languages, which perform the chosen complexity measurements.
The results of these tools are extracted, transformed into a common data model and saved in a typical \emph{Extract, Transform, Load} (ETL) process, allowing analyses on well defined data structures.
Analyzr aggregates the analysis results and presents visualizations to the user in a web interface, see Figure~\ref{fig:metrics}.

\subsection{Architecture}
\begin{figure}
    \centering
    \includegraphics[width=0.85\columnwidth]{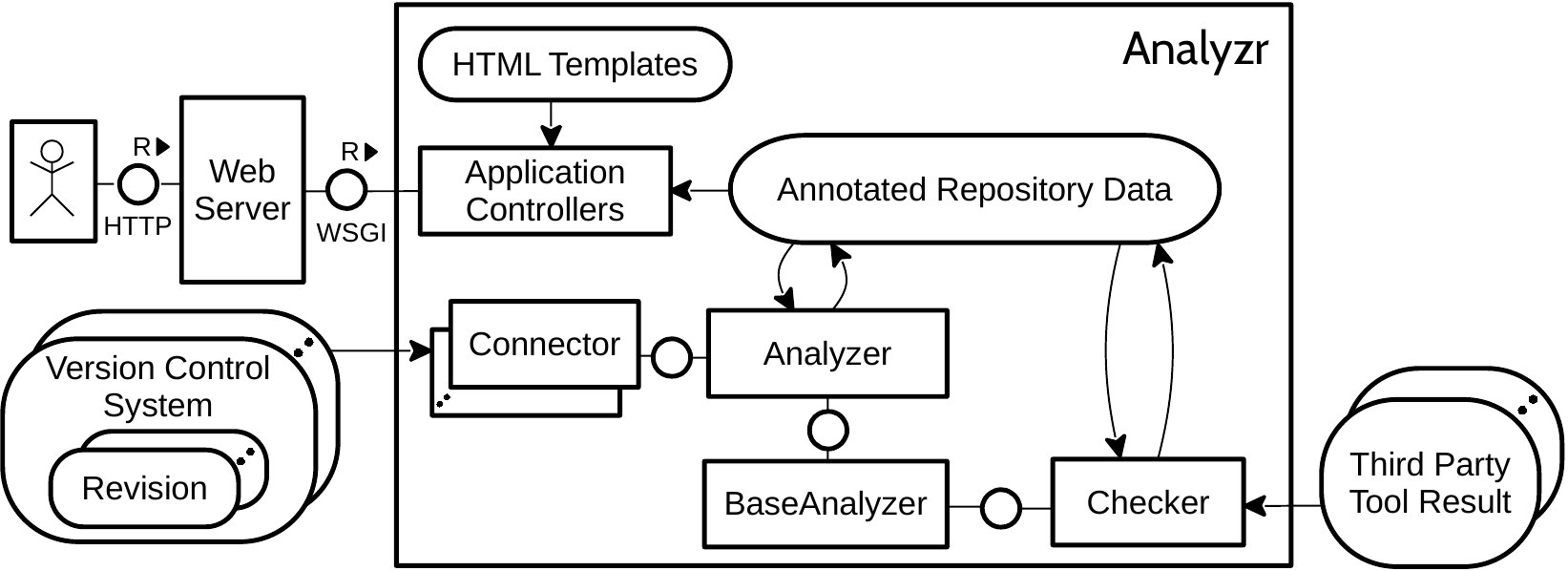}
    \caption{FMC block diagram of the back end architecture of Analyzr.}
    \label{fig:backend-architecture}
\end{figure}

Analyzr itself is split into a back end, using the Python web-framework Django~\cite{django}, and a front end, using HTML5 and JavaScript, which is accessed with a browser.
This allows long-running analyses tasks to be performed on the server and not strain client resources.
Figure~\ref{fig:backend-architecture} depicts the back end, which collects data from the different repositories that are to be analysed and stores the data which is produced during analyses.
It exposes a REST interface to the front end, which presents a user interface to explore the data.

\subsection{Data Model}
\begin{figure}
    \centering
    \includegraphics[width=0.9\columnwidth]{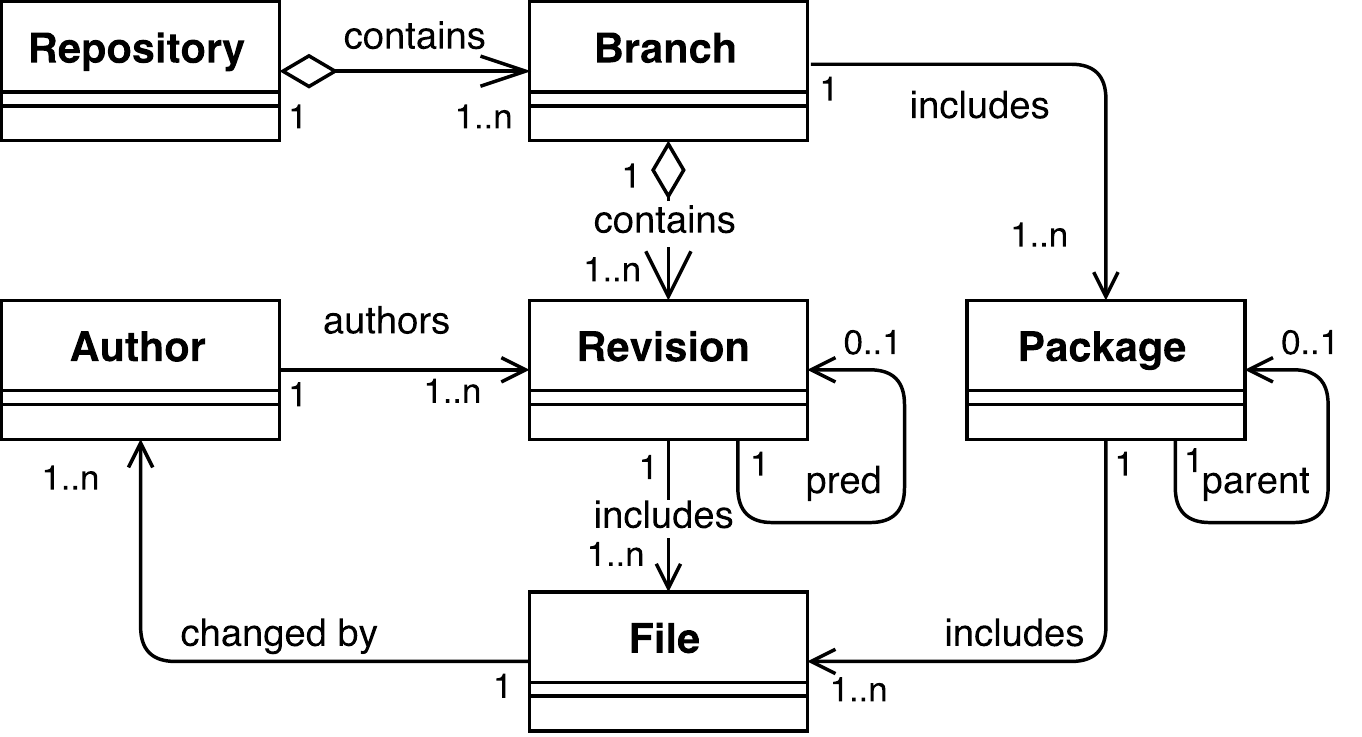}
    \caption{UML class diagram of the models used to store gathered repository information.}
    \label{fig:data-model}
\end{figure}

The employed data model reflects the basic structure of a software repository.
It is shown in Figure~\ref{fig:data-model}.
A \emph{Repo} entity holds information such as the location of the remote repository and user credentials and has a number of branches, which in turn are connected to a number of revisions, e.g. commits.
For the sake of query performance and join avoidance, some information, such as the revision author is kept redundantly, in both the file and revision entities.

\subsection{Extensibility}
\begin{figure}
    \centering
    \includegraphics[width=0.9\columnwidth]{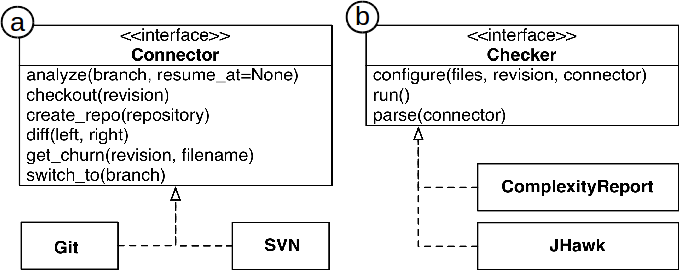}
    \caption{The interfaces which have to be implemented when adding a new \emph{Connector} or \emph{Checker} to Analyzr.}
    \label{fig:interfaces}
\end{figure}

The task of communicating with the version control system is handled by \emph{Connector} implementations.
Currently, Subversion and Git repositories are supported.
Every specialised connector is able to extract information from each revision in a given branch.
Other version control systems can be added by implementing the minimal interface shown in Figure~\ref{fig:interfaces}a.
\emph{Checker} classes (see Figure~\ref{fig:interfaces}b) wrap third party code analysis tools, ensuring that they expose a common interface.
Multiple checker instances can be used for analysing a certain programming language, so that that weaknesses in the analysis of one tool can be covered by another.

\subsection{Third Party Tools}
Our approach relies on proven, time-tested third party tools, that implement the described code metric algorithms.
For Java as well as JavaScript, specialized tools were chosen.

\subsubsection{JHawk}
Java, as a statically typed language, allows computing a variety of metrics, ranging from generic complexity to object oriented ones~\cite{harrison1997overview}.
JHawk~\cite{jhawk} offers support for all of the software metrics required by Analyzr.
It is possible to start measurements in a given directory and restrict the set of files which will be analysed.
Using this mechanism we could incorporate the knowledge available in the version control system to only reassess changed files.
JHawk produces an XML that is then further processed and loaded into the database.

\subsubsection{Complexity Report}
Complexity Report~\cite{complexityreport} is an open source JavaScript software, which can be run from the command line. 
As JavaScript is loosely typed, it does not natively facilitate the concept of packages, therefore $C_e$ and $C_a$ cannot be measured. 
Nevertheless, we are able to measure the McCabe complexity, the Halstead metrics, and the source lines of code (SLOC) using Complexity Report.

\section{Metric Aggregation and Expertise Extraction}
One of the main functions of Analyzr is to aggregate the different collected complexity measures into a single score, representing developer expertise for components.
Metrics are computed on a file basis, instead of directly at package or component level, as this allows more fine-grained analyses and supports applying weighting operations earlier in the calculation.
These values are then again aggregated on a directory level to form overall metrics for packages.
However, some of the employed complexity measurements produce results on a function level. 
These low-level results need to be aggregated into a single value.
Regular aggregation methods, such as mean and median are not suitable for all code metrics~\cite{vasilescu2011you}.
As code metrics have their own domain and semantics they are hard to compare.
This is an issue Mordal et al. tried to solve with their \emph{Squale} model~\cite{mordal2012software}.

\subsection{Software Quality Enhancement (Squale)}
\label{sec:squale}
Squale summarizes metric values by highlighting those which are problematic and weighs those values more that have recently improved~\cite{mordal2012software}, e.g. it emphasizes improvements in badly-rated system parts.
In general, Squale provides a bounded, continuous scale for the comparison of metric values.
It combines low-level marks into individual marks, which are then aggregated to a global mark:
\begin{enumerate}
    \item \textbf{Low-level marks} are raw values retrieved from source code analysis, either manual metrics, assessed by humans, or automated tools, such as code metrics or rule checking. 
    \item \textbf{Individual marks} are computed from low-level marks.
    The thresholds for \enquote{good} or \enquote{bad} values with respect to project quality are determined by experts for the given field.
    They are mapped to a unified scale from 0 to 3 to allow comparisons.
\end{enumerate}

After each individual mark (IM) has been computed they are aggregated using a weighting function.
It is is defined as:
\begin{equation*}
    weight(\text{IM}) = \lambda^{-\text{IM}}
\end{equation*}
$\lambda$ defines the strength of the weighting. Common strength values are 3, 9, and 30 for soft, medium, and hard, respectively.
Hard weightings give more weight to bad results than soft weightings.
The global mark $\text{GM}^{\lambda}$, combining all individual marks~\cite{mordal2012software}, is computed as:
\begin{equation*}
\text{GM}^{\lambda} = - \log_{\lambda}{\left(\frac{1}{\left\vert{IM}\right\vert} \cdot \sum^{\left\vert{IM}\right\vert}_{i = 1}{weight(\text{IM}_i)}\right)}
\end{equation*}

\subsection{Computation of Employed Metrics}
\label{sec:employed}
To compute the individual marks for the employed metrics presented in Section~\ref{sec:complexitymeasures} we used the formulas presented in Table~\ref{tab:mark-overview}, originally developed by Balmas et al.~\cite{Balmas:2010dq}.
The \emph{lower threshold} and \emph{upper threshold} describe the boundaries above and below which constant values of 0 and 3 are returned.
If the raw metric value is lower than the lower threshold, the code is assumed to not be complex and an individual mark of 3 is returned, indicating low relevance for refactoring.
In the case of a raw metric value being larger than the upper threshold, an individual mark of 0 is returned, indicating a strong need for further review.
Individual marks for raw metric values between the two thresholds are computed using the presented formulas in Table~\ref{tab:mark-overview}.

\begin{table}
    \caption{Individual mark computation for the complexity metrics employed by Analyzr.}
    \label{tab:mark-overview}
    \begin{center}
        The lower and upper thresholds describe those input values below and above which constants are returned for $\text{IM}_{i}$. 
        
        \vspace{0.2cm}
        
        \begin{tabularx}{\columnwidth}{|p{1.35cm}|p{3.4cm}|p{0.7cm}|p{0.7cm}|l|}
            \hline
            \textbf{Metric} & \textbf{Formula} & \textbf{Lower Thr.} & \textbf{Upper Thr.} & \textbf{Ref.}\\ \hline
            Cyclomatic Complexity & $\text{IM}_{cc} =$ \scalebox{1.4}{$2^{\slfrac{\left(7 - cc\right)}{3.5}}$} & 2 & 19 & \cite{Balmas:2010dq} \\ \hline
            Halstead Volume & $\text{IM}_{hv} =$ \scalebox{1.4}{$3 - \frac{3 \times hv}{1000}$} & 20 & 1000 & \cite{Serebrenik:2011qf} \\ \hline
            Halstead Difficulty & $\text{IM}_{hd} =$ \scalebox{1.4}{$3 - \frac{3 \times hd}{50}$} & 10 & 50 & \cite{Serebrenik:2011qf} \\ \hline
            Afferent Coupling & $\text{IM}_{C_a} =$ \scalebox{1.4}{$2^{\slfrac{\left(30 - C_a\right)}{7}}$} & 19 & 60 & \cite{Balmas:2010dq}\\ \hline
            Efferent Coupling & $\text{IM}_{C_e} =$ \scalebox{1.4}{$2^{\slfrac{\left(10 - C_e\right)}{2}}$} & 6 & 19 & \cite{Balmas:2010dq} \\ \hline
        \end{tabularx}
    \end{center}
\end{table}

\subsection{Extracting Expertise of Developers}
Our main goal is to determine the expertise of individual developers for specific program parts.
For that purpose, we only regard developers having committed to the respective parts within a certain time frame\footnote{In our study we picked 62 days, as this served as a distinguishing factor between temporary and permanent leave, in the context of the studied company.}. 
Within this time frame, the set of revisions ($r$) created by an author ($a$), from the set of all authors ($A$) up to a certain point in time ($T$) is defined as:
\begin{equation*}
R_T(a) = \{ r \mid r \in R_T \wedge a \in A \wedge \mathrm{author}(r) = a \}
\end{equation*}

Naive approaches like counting the number of commits or the number of lines changed as an indicator for expertise can be used for a first estimation.
However, we are confident that for developers in an established team the expertise concerning a program part is better reflected in the quality of the code they produce.
As most development takes place on an already existing code base, we focus on the changes of the quality metrics, i.e. deltas, in code metrics.
For each revision within the set $R_T(a)$, we check whether the global mark, computed using Squale, improved or deteriorated.
These changes are attributed to the developer who authored the commit.
We count all commits resulting in increases and decreases of software quality and calculate the developer's individual score ($\mathrm{score}_T(a)$):

\begin{equation*}
qi = min\left(\frac{\mathrm{increases}_T(a)}{\mathrm{decreases}_T(a)},1\right)
\end{equation*}
\begin{equation*}
\mathrm{score}_T(a) = qi \times \ln \left( 1 + | R_T(a) | \right)\\
\end{equation*}

\noindent
The ratio of commits that increase versus those that decrease software quality ($qi$, quality impact) is multiplied with the logarithmic smoothed amount of total commits of the author.
The formula incorporates the total amount of commits, but does not double the score for double the number of commits, which would skew the score in favor of long-serving developers.
In the unlikely case that a developer has not produced a commit which decreased the code quality, we fall back to evaluating the number of commits that were produced, as $qi$ is 1.
However, the normal case in software development is that there are more commits which decrease, rather than increase code quality~\cite{mens2004survey}, resulting in $qi$ being lower than 1.

The score therefore combines the experience of a developer, expressed by the number of his commits with his quality impact, to reflect his expertise.

\section{Test Bed Selection}
In order to evaluate the quality of Analyzr results, the software was deployed at a collaborating software company and real development data was analyzed.
The company chosen was a German software development company in the sector of business process modelling.
It was chosen as one of the authors was employed there and had insights into the structures and processes within the company.

\subsection{Company Introduction}
As the company that was analyzed builds mostly applications for the web, they practice a separation between front and back end code.
Of the then about thirty developers in the company, around ten are active in the front end and the other twenty are concerned with back end development.
The back end of the software is implemented in Java, whereas the front end is a JavaScript application.
Furthermore, the developed software is split into well-defined components, providing clearly bound domains of expertise.
Currently code metrics are not an official part of the development process of the company.
However, code reviews are performed amongst developers, with the aim of keeping the code maintainable.

\subsection{Time Frame}
When attempting to determine the current experts of a software project the history of the project that is being analyzed needs to be restricted to allow focusing on those developers that have recently contributed code and are well-versed with the current status of the software.
For the studied company, our observations and interviews pointed to an optimal time frame for analysis of around 62 days.
This period is long enough so developers who are on a temporal leave do not get sorted out, but it is short enough to exclude developers who are no longer active in a certain module or have left the company.

\subsection{Threats to Validity}
Real world development data from a company has inconsistencies, that make analysis challenging.
In the data set collected for this paper, credentials of developers differed over time, for example if a developer changed his user name in the version control system.
We consolidated these changes in a manual correction phase.
Data was gathered from the \emph{master} (Git) and \emph{trunk} (SVN) branches of the version control systems, as only this code represents usable increments of the software.
Code that was in other, possibly specialized development or feature branches was not analyzed.

\section{Evaluation}
In order to evaluate the results of Analyzr and compare them to the expectations of the developers, a survey was devised.
Survey participants were developers who volunteered.
Overall, 13 developers participated, aged 20 to 30 years.
Among them were 12 males and 1 female; 5 participants were seniors, while 8 were junior developers.
For the following analysis all names were anonymized.
The survey consisted of two main parts:
\begin{itemize}
    \item \textbf{Expert Selection} Developers self-assessed whether they were active in the front or the back end.
    Based on their decision they were presented with the respective components of their division and were asked to select an expert for each. They were free to name two other qualified developers as well.
    \item \textbf{Proposal Evaluation} Developers were presented with the top three component experts identified by Analyzr and were asked to rate the accuracy of each result.
\end{itemize}

\subsection{Expert Selection}
\label{sec:expertselection}
Table~\ref{tbl:experts} summarizes the selections of the company's staff for first, second and third choice of rank of expertize for each software component.
The percentages given represent the consensus of the developers for the given rank, meaning that for example Marler was chosen most often as the best expert for Analytics (representing 60\% of the votes), and the missing 40\% of the vote-shares for the first rank favored other candidates.
A developer could be nominated as first, second and third choice at the same time, if respondents see someone as the only choice (or best choice and last fallback, if the second best expert is not available).
For 6 components out of 11 in the front end and 2 components out of 6 in the back end all surveyed developers agreed on the first choice of expert.
High uncertainty among developers concerning expert selection, i.e. less than 50\% agreement, was only apparent in 6 out of 51 choices.
This data lends credibility to the hypothesis that developers have a specific component expert in mind, whom they would most likely consult if they had a question in that specific domain.
Furthermore, we found a high level of agreement who was considered to be among the top experts when asking the developers.
However, both in the front as well as in the back end, only two distinct developers were voted as first choice: Marler and Moyer as well as Marston And Anstine, respectively.
While this is certainly flattering for these developers, it also introduces problems.
They will most likely be interrupted frequently and if they are out of the office, developer's single point of contact in case of questions is lost.

\begin{table}
    \begin{center}
        FRONT END COMPONENTS
        
        \vspace{0.2cm}
        
        \begin{tabular}{|l|l|l|l|}
            \hline
            \textbf{Component} & \textbf{1st choice} & \textbf{2nd choice} & \textbf{3rd choice} \\ \hline
            Administration & Marler (100\%) & Prud (100\%) & Braaten (100\%) \\ \hline
            Analytics & Marler (60\%) & Braaten (100\%) & Marler (100\%) \\ \hline
            Comparator & Marler (100\%) & Waring (100\%) & Marcuso (50\%) \\ \hline
            Editor & Marler (100\%) & Moyer (60\%) & Mayberry (67\%) \\ \hline
            Explorer & Marler (100\%) & Waring (50\%) & Mayberry (67\%) \\ \hline
            Glossary & Marler (100\%) & Waring (100\%) & Moyer (67\%) \\ \hline
            Utils & Moyer (60\%) & Mayberry (50\%) & Waring (100\%) \\ \hline
            Portal & Marler (100\%) & Waring (100\%) & Moyer (100\%) \\ \hline
            Quick Model & Marler (60\%) & Waring (100\%) & Marler (50\%) \\ \hline
            Simulation & Marler (40\%) & Braaten (50\%) & Marler (100\%) \\ \hline
            Testing & Moyer (80\%) & Salmeron (50\%) & Marler (25\%) \\ \hline
        \end{tabular}
        
        \vspace{0.4cm}
        
        BACK END COMPONENTS
        
        \vspace{0.2cm}
        
        \begin{tabular}{|l|l|l|l|}
            \hline
            \textbf{Component} & \textbf{1st choice} & \textbf{2nd choice} & \textbf{3rd choice} \\ \hline
            Diagram API & Marston (88\%) & Gillette (67\%) & Prouty (50\%) \\ \hline
            Glossary & Anstine (38\%) & Marston (57\%) & Gillette (33\%) \\ \hline
            Platform & Anstine (100\%) & Prud (57\%) & Marston (50\%) \\ \hline
            SVG Renderer & Marston (88\%) & Gillette (50\%) & Braaten (50\%) \\ \hline
            User Mgmt & Anstine (88\%) & Prud (50\%) & Braaten (40\%) \\ \hline
            Warehouse & Anstine (100\%) & Prud (75\%) & Braaten (33\%) \\ \hline
        \end{tabular}
    \end{center}
    \caption{Experts for the different front end (top) and back end (bottom) components as voted by company developers. Displayed percentages represent the consensus of developers for a given rank.}
    \label{tbl:experts}
\end{table}

\begin{figure*}[htb]
    \centering
    \includegraphics[width=\textwidth]{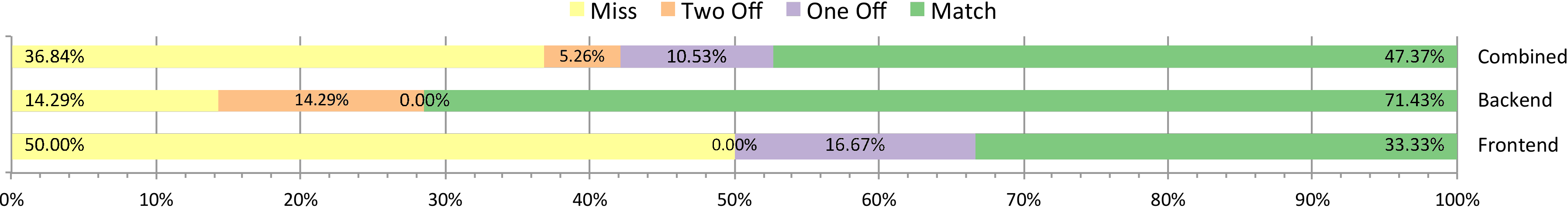}
    \caption{Accuracy of our findings considering only the first choice compared to the statements made by the staff. The results are grouped into perfect match, one off, two off, and miss. The combined percentages are skewed towards the front end, as there are more front than back end components.}
    \label{fig:first-choice-matches}
\end{figure*}

The accuracy of Analyzr results was evaluated by comparing the level of agreement between the lists of manually and automatically selected component experts.
The developers, who chose the experts manually, did not have prior knowledge of Analyzr results.
Figure~\ref{fig:first-choice-matches} summarizes the accuracy of Analyzr predictions for the first choice of component experts for the whole company repository.
If the prediction equalled the manually identified expert, it was classified as a match.
Otherwise an Anaylzr result can be classified one or two off, depending on whether the prediction missed the top result by one or two ranks.
If the prediction was not included in the list of manually selected component experts it was considered a miss.
Analyzr was able to produce a match in 47,37\% of observed cases, with 15,79\% of cases classified as one or two ranks off.
This is influenced by the fact that in the front end in 50\% of cases the manually identified expert was missed.
This raises the question whether some of the developers simply did not know who the component experts were and misidentified them, or whether Analyzr results were inaccurate.
In order to answer this question the expert proposals by Analyzr were directly rated by the developers.

\begin{figure*}[hbt]
    \centering
    \includegraphics[width=\textwidth]{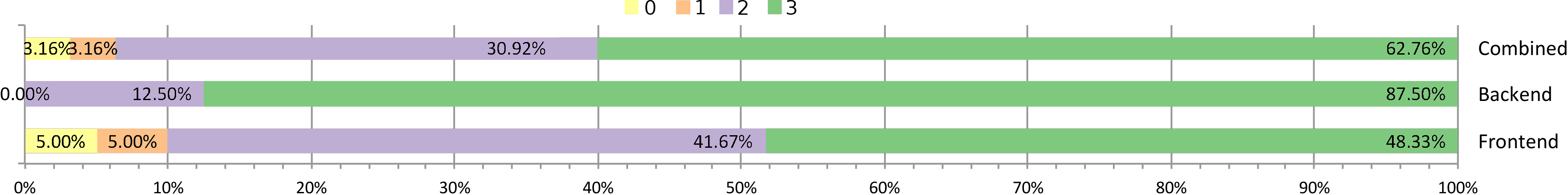}
    \caption{Acceptance of the computed experts, regarding only the first choice by the staff. The numeric values represent: disagreement (0), weak disagreement (1), weak agreement (2), agreement (3). The combined percentages are skewed towards the front end, as there are more front than back end components.}
    \label{fig:first-choice-acceptance}
\end{figure*}

\subsection{Proposal Evaluation}
For each of the three experts identified by Analyzr for the software components, all interviewed developers were asked to rate the results on a 4-point scale: \enquote{strongly disagree} (0), \enquote{disagree} (1), \enquote{agree} (2), \enquote{strongly agree} (3).
The choice of \enquote{neutral} was deliberately omitted, as it has no semantics concerning the acceptance of a proposed expert.
Considering the evaluated expertise of a developer, this means the proposed developer ...
\begin{itemize}
    \item[\textbf{0}] ... does not have knowledge of the component. 
    \item[\textbf{1}] ... does not have enough knowledge to be considered a component expert.
    \item[\textbf{2}] ... is acceptable as component expert, but someone else would be better suited.
    \item[\textbf{3}] ... is identified correctly as component expert and the result is considered useful.
\end{itemize}
The ratings regarding the first choice of expert presented by Analyzr are summarized in Figure~\ref{fig:first-choice-acceptance}.
Overall, experts recommended by Analyzr as the first choice were rated as either acceptable or completely correct (2 or 3 points) with a rate of 100\% in the back end and 90\% in the front end.
This is evidence for the hypothesis that there are non-obvious component experts that are unknown to developers and cannot be identified by simply asking staff members.
However, Analyzr was able to identify these experts who were accepted by the vast majority of survey participants.

\section{Conclusion and Future Work}
In this paper we presented Analyzr, a framework for eliciting the expertise of developers for software components using complexity analysis on source code.
We showed that using our approach, it is feasible to extract component experts in a case study with a medium-size software development company.
The experts we found and suggested differed from those that developers picked intuitively.
However, the algorithmically extracted experts were rated by developers as accurate in the vast majority of cases.
This lends credibility to the hypothesis that Analyzr was able to find \enquote{hidden experts}, i.e. developers who have a lot of specific component knowledge and can answer questions, but may not be the intuitive choice.
Identifying these experts is helpful for organisations as no longer all questions and enquiries have to be directed to the single obvious expert.
This relieves pressure from this expert and enables wider distribution of knowledge as more developers become involved in solving challenging questions.
Furthermore, developers can visualize and track their statistics, see the progress they are making, which provides an additional incentive to produce cleaner code.
Future work will focus on iteratively improving the software based on user feedback after having productively used it at the surveyed company for an extended period of time.
Especially, improvements to metric selection and thresholds will further enhance the accuracy of results.
For the case of analysing object-oriented languages, more specific metrics, such as the Method Hiding Factor or the Attribute Hiding Factor, could be employed~\cite{harrison1997overview}.
These can provide better insights into the code base at the cost of general applicability.


%
\bibliography{references}

\begin{thebibliography}{10}
\providecommand{\url}[1]{#1}
\csname url@samestyle\endcsname
\providecommand{\newblock}{\relax}
\providecommand{\bibinfo}[2]{#2}
\providecommand{\BIBentrySTDinterwordspacing}{\spaceskip=0pt\relax}
\providecommand{\BIBentryALTinterwordstretchfactor}{4}
\providecommand{\BIBentryALTinterwordspacing}{\spaceskip=\fontdimen2\font plus
\BIBentryALTinterwordstretchfactor\fontdimen3\font minus
  \fontdimen4\font\relax}
\providecommand{\BIBforeignlanguage}[2]{{%
\expandafter\ifx\csname l@#1\endcsname\relax
\typeout{** WARNING: IEEEtran.bst: No hyphenation pattern has been}%
\typeout{** loaded for the language `#1'. Using the pattern for}%
\typeout{** the default language instead.}%
\else
\language=\csname l@#1\endcsname
\fi
#2}}
\providecommand{\BIBdecl}{\relax}
\BIBdecl

\bibitem{williams_pair_2002}
L.~Williams and R.~Kessler, \emph{Pair Programming Illuminated}.\hskip 1em plus
  0.5em minus 0.4em\relax Boston, MA, USA: Addison-Wesley Longman Publishing
  Co., Inc., 2002.

\bibitem{nordberg03}
M.~E. Nordberg~III, ``Managing code ownership,'' \emph{IEEE Softw.}, vol.~20,
  no.~2, pp. 26--33, Mar. 2003.

\bibitem{Ehrlich:2006:LEG:1170744.1171374}
K.~Ehrlich and K.~Chang, ``Leveraging expertise in global software teams: Going
  outside boundaries,'' in \emph{Proceedings of the IEEE International
  Conference on Global Software Engineering}, ser. ICGSE '06.\hskip 1em plus
  0.5em minus 0.4em\relax Washington, DC, USA: IEEE Computer Society, 2006, pp.
  149--158.

\bibitem{Faraj:2000:CES:970301.970322}
S.~Faraj and L.~Sproull, ``Coordinating expertise in software development
  teams,'' \emph{Manage. Sci.}, vol.~46, no.~12, pp. 1554--1568, Dec. 2000.

\bibitem{herbsleb1999beyondconwayslaw}
J.~Herbsleb and R.~Grinter, ``Architectures, coordination, and distance:
  Conway's law and beyond,'' \emph{Software, IEEE}, vol.~16, no.~5, pp. 63
  --70, sep/oct 1999.

\bibitem{Beck1999}
K.~Beck, \emph{{Extreme Programming Explained: Embrace Change}}.\hskip 1em plus
  0.5em minus 0.4em\relax Addison-Wesley Professional, 2000.

\bibitem{conway}
M.~E. Conway, ``How do committees invent,'' \emph{Datamation}, vol.~14, no.~4,
  pp. 28--31, 1968.

\bibitem{agilealliance15}
\BIBentryALTinterwordspacing
{Agile Alliance}, ``Agile glossary --- collective ownership,'' 2015, [accessed
  4-August-2016]. [Online]. Available:
  \url{https://www.agilealliance.org/glossary/collective-ownership/}
\BIBentrySTDinterwordspacing

\bibitem{coleman1994using}
D.~Coleman, D.~Ash, B.~Lowther, and P.~Oman, ``Using metrics to evaluate
  software system maintainability,'' \emph{Computer}, vol.~27, no.~8, pp.
  44--49, 1994.

\bibitem{clark2008measuring}
M.~Clark, B.~Salesky, C.~Urmson, and D.~Brenneman, ``Measuring software
  complexity to target risky modules in autonomous vehicle systems,'' in
  \emph{Proceedings of the AUVSI North America Conference}, 2008.

\bibitem{watson1996structured}
A.~H. Watson, T.~J. McCabe, and D.~R. Wallace, ``Structured testing: A testing
  methodology using the cyclomatic complexity metric,'' \emph{NIST special
  Publication}, vol. 500, no. 235, pp. 1--114, 1996.

\bibitem{nagappan2006mining}
N.~Nagappan, T.~Ball, and A.~Zeller, ``Mining metrics to predict component
  failures,'' in \emph{Proceedings of the 28th international conference on
  Software engineering}.\hskip 1em plus 0.5em minus 0.4em\relax ACM, 2006, pp.
  452--461.

\bibitem{vasilescu2011no}
B.~Vasilescu, A.~Serebrenik, and M.~van~den Brand, ``By no means: A study on
  aggregating software metrics,'' in \emph{Proceedings of the 2nd International
  Workshop on Emerging Trends in Software Metrics}.\hskip 1em plus 0.5em minus
  0.4em\relax ACM, 2011, pp. 23--26.

\bibitem{vasilescu2011you}
------, ``You can't control the unfamiliar: A study on the relations between
  aggregation techniques for software metrics,'' in \emph{2011 27th IEEE
  International Conference on Software Maintenance (ICSM)}, Sept 2011, pp.
  313--322.

\bibitem{mordal2012software}
K.~Mordal, N.~Anquetil, J.~Laval, A.~Serebrenik, B.~Vasilescu, and S.~Ducasse,
  ``Software quality metrics aggregation in industry,'' \emph{Journal of
  Software: Evolution and Process}, 2012.

\bibitem{Avelino2016}
\BIBentryALTinterwordspacing
G.~Avelino, L.~Passos, A.~Hora, and M.~T. Valente, ``{A novel approach for
  estimating Truck Factors},'' in \emph{2016 IEEE 24th International Conference
  on Program Comprehension (ICPC)}, vol. 2016-July, no. Dcc.\hskip 1em plus
  0.5em minus 0.4em\relax IEEE, may 2016, pp. 1--10. [Online]. Available:
  \url{http://ieeexplore.ieee.org/document/7503718/}
\BIBentrySTDinterwordspacing

\bibitem{Bird2011}
C.~Bird, N.~Nagappan, B.~Murphy, H.~Gall, and P.~Devanbu, ``{Don't Touch My
  Code!: Examining the Effects of Ownership on Software Quality},''
  \emph{Proceedings of the 19th ACM SIGSOFT Symposium and the 13th European
  Conference on Foundations of Software Engineering}, pp. 4--14, 2011.

\bibitem{Foucault2014}
\BIBentryALTinterwordspacing
M.~Foucault, J.-R. Falleri, and X.~Blanc, ``{Code ownership in open-source
  software},'' in \emph{Proceedings of the 18th International Conference on
  Evaluation and Assessment in Software Engineering - EASE '14}.\hskip 1em plus
  0.5em minus 0.4em\relax New York, New York, USA: ACM Press, 2014, pp. 1--9.
  [Online]. Available:
  \url{http://dl.acm.org/citation.cfm?doid=2601248.2601283}
\BIBentrySTDinterwordspacing

\bibitem{Thongtanunam2016}
\BIBentryALTinterwordspacing
P.~Thongtanunam, S.~McIntosh, A.~E. Hassan, and H.~Iida, ``{Revisiting code
  ownership and its relationship with software quality in the scope of modern
  code review},'' in \emph{Proceedings of the 38th International Conference on
  Software Engineering - ICSE '16}, no.~1.\hskip 1em plus 0.5em minus
  0.4em\relax New York, New York, USA: ACM Press, 2016, pp. 1039--1050.
  [Online]. Available:
  \url{http://dl.acm.org/citation.cfm?doid=2884781.2884852}
\BIBentrySTDinterwordspacing

\bibitem{McDonald2000}
\BIBentryALTinterwordspacing
D.~W. McDonald and M.~S. Ackerman, ``{Expertise Recommender: A Flexible
  Recommendation System and Architecture},'' in \emph{Proceedings of the 2000
  ACM conference on Computer supported cooperative work - CSCW '00}.\hskip 1em
  plus 0.5em minus 0.4em\relax New York, New York, USA: ACM Press, 2000, pp.
  231--240. [Online]. Available:
  \url{http://portal.acm.org/citation.cfm?doid=358916.358994}
\BIBentrySTDinterwordspacing

\bibitem{schuler2008mining}
D.~Schuler and T.~Zimmermann, ``Mining usage expertise from version archives,''
  in \emph{Proceedings of the 2008 international working conference on Mining
  software repositories}.\hskip 1em plus 0.5em minus 0.4em\relax ACM, 2008, pp.
  121--124.

\bibitem{Anvik:2006}
\BIBentryALTinterwordspacing
J.~Anvik, L.~Hiew, and G.~C. Murphy, ``Who should fix this bug?'' in
  \emph{Proceedings of the 28th International Conference on Software
  Engineering}, ser. ICSE '06.\hskip 1em plus 0.5em minus 0.4em\relax New York,
  NY, USA: ACM, 2006, pp. 361--370. [Online]. Available:
  \url{http://doi.acm.org/10.1145/1134285.1134336}
\BIBentrySTDinterwordspacing

\bibitem{Tian2016}
\BIBentryALTinterwordspacing
Y.~Tian, D.~Wijedasa, D.~Lo, and C.~{Le Goues}, ``{Learning to rank for bug
  report assignee recommendation},'' in \emph{2016 IEEE 24th International
  Conference on Program Comprehension (ICPC)}, vol. 2016-July.\hskip 1em plus
  0.5em minus 0.4em\relax IEEE, may 2016, pp. 1--10. [Online]. Available:
  \url{http://ieeexplore.ieee.org/document/7503715/}
\BIBentrySTDinterwordspacing

\bibitem{Venkataramani:2013}
\BIBentryALTinterwordspacing
R.~Venkataramani, A.~Gupta, A.~Asadullah, B.~Muddu, and V.~Bhat, ``Discovery of
  technical expertise from open source code repositories,'' in
  \emph{Proceedings of the 22Nd International Conference on World Wide Web},
  ser. WWW '13 Companion.\hskip 1em plus 0.5em minus 0.4em\relax New York, NY,
  USA: ACM, 2013, pp. 97--98. [Online]. Available:
  \url{http://doi.acm.org/10.1145/2487788.2487832}
\BIBentrySTDinterwordspacing

\bibitem{latoza2006maintaining}
T.~D. LaToza, G.~Venolia, and R.~DeLine, ``Maintaining mental models: a study
  of developer work habits,'' in \emph{Proceedings of the 28th international
  conference on Software engineering}.\hskip 1em plus 0.5em minus 0.4em\relax
  ACM, 2006, pp. 492--501.

\bibitem{hoffman2000darker}
D.~Hoffman, ``The darker side of metrics,'' in \emph{Pacific Northwest Software
  Quality Conference, Portland, Oregon}, 2000.

\bibitem{riaz2009systematic}
M.~Riaz, E.~Mendes, and E.~Tempero, ``A systematic review of software
  maintainability prediction and metrics,'' in \emph{Proceedings of the 2009
  3rd International Symposium on Empirical Software Engineering and
  Measurement}.\hskip 1em plus 0.5em minus 0.4em\relax IEEE Computer Society,
  2009, pp. 367--377.

\bibitem{german2003software}
A.~German, ``Software static code analysis lessons learned,'' \emph{Crosstalk},
  vol.~16, no.~11, 2003.

\bibitem{kaner2004software}
C.~Kaner and W.~P. Bond, ``Software engineering metrics: What do they measure
  and how do we know?'' \emph{methodology}, vol.~8, p.~6, 2004.

\bibitem{basili1996validation}
V.~R. Basili, L.~C. Briand, and W.~L. Melo, ``A validation of object-oriented
  design metrics as quality indicators,'' \emph{Software Engineering, IEEE
  Transactions on}, vol.~22, no.~10, pp. 751--761, 1996.

\bibitem{graves2000predicting}
T.~L. Graves, A.~F. Karr, J.~S. Marron, and H.~Siy, ``Predicting fault
  incidence using software change history,'' \emph{Software Engineering, IEEE
  Transactions on}, vol.~26, no.~7, pp. 653--661, 2000.

\bibitem{mccabe1976complexity}
T.~J. McCabe, ``A complexity measure,'' \emph{Software Engineering, IEEE
  Transactions on}, no.~4, pp. 308--320, 1976.

\bibitem{halstead1977potential}
M.~Halstead, ``Potential impacts of software science on software life cycle
  management,'' 1977.

\bibitem{shen1983software}
V.~Y. Shen, S.~D. Conte, and H.~E. Dunsmore, ``Software science revisited: A
  critical analysis of the theory and its empirical support,'' \emph{Software
  Engineering, IEEE Transactions on}, no.~2, pp. 155--165, 1983.

\bibitem{henry1981software}
S.~Henry and D.~Kafura, ``Software structure metrics based on information
  flow,'' \emph{Software Engineering, IEEE Transactions on}, no.~5, pp.
  510--518, 1981.

\bibitem{crockford2008javascript}
D.~Crockford, \emph{JavaScript: the good parts}.\hskip 1em plus 0.5em minus
  0.4em\relax O'Reilly Media, Inc., 2008.

\bibitem{redin2008use}
R.~M. Redin, M.~F. Oliveira, L.~B. Brisolara, J.~C. Mattos, L.~C. Lamb, F.~R.
  Wagner, and L.~Carro, ``On the use of software quality metrics to improve
  physical properties of embedded systems,'' in \emph{Distributed Embedded
  Systems: Design, Middleware and Resources}.\hskip 1em plus 0.5em minus
  0.4em\relax Springer, 2008, pp. 101--110.

\bibitem{django}
{Django Software Foundation}, ``Django web-framework,''
  \url{https://www.djangoproject.com/}.

\bibitem{harrison1997overview}
R.~Harrison, S.~Counsell, and R.~Nithi, ``An overview of object-oriented design
  metrics,'' in \emph{Software Technology and Engineering Practice, 1997.
  Proceedings., Eighth IEEE International Workshop on [incorporating Computer
  Aided Software Engineering]}.\hskip 1em plus 0.5em minus 0.4em\relax IEEE,
  1997, pp. 230--235.

\bibitem{jhawk}
\BIBentryALTinterwordspacing
{Virtual Machinery}, ``Jhawk product overview,'' 2015, [accessed
  4-August-2016]. [Online]. Available:
  \url{http://www.virtualmachinery.com/jhawkprod.htm}
\BIBentrySTDinterwordspacing

\bibitem{complexityreport}
\BIBentryALTinterwordspacing
escomplex, ``complexity-report,'' 2016. [Online]. Available:
  \url{https://github.com/escomplex/complexity-report}
\BIBentrySTDinterwordspacing

\bibitem{Balmas:2010dq}
F.~Balmas, F.~Bellingrad, F.~Denier, S.~Ducasse, B.~Franchet, J.~Laval,
  K.~Mordal-Manet, and P.~Vaillergues, ``The squale quality model. mod{\`e}le
  enrichi d'agr{\'e}gation des pratiques pour java et c++,'' INRIA, Tech. Rep.,
  2010.

\bibitem{Serebrenik:2011qf}
A.~Serebrenik, ``2is55 software evolution,'' 2011.

\bibitem{mens2004survey}
T.~Mens and T.~Tourw{\'e}, ``A survey of software refactoring,'' \emph{IEEE
  Transactions on software engineering}, vol.~30, no.~2, pp. 126--139, 2004.

\end{thebibliography}
\bibliographystyle{IEEEtran}

\end{document}